\documentclass[twocolumn]{aastex7}

\usepackage{siunitx} 
\usepackage{amsmath} 
\usepackage[utf8]{inputenc}
\usepackage{float}  

\usepackage{booktabs}
\usepackage{makecell}
\usepackage{multirow}

\usepackage[flushleft]{threeparttable}

\usepackage{caption}

\usepackage{array}
\usepackage{graphicx}
\usepackage{threeparttable}
\usepackage{adjustbox}
\usepackage{ragged2e}
\usepackage{etoolbox}

\usepackage{fixltx2e} 

\begin{document}

\title{New symbiotic stars or candidates in LAMOST low resolution spectra}

\author{Yabing Zhao}
\affiliation{School of Physical Science and Technology, Xinjiang University, Urumqi 830046, China}
\email{107552300789@stu.xju.edu.cn}  

\author{Sufen Guo*}
\affiliation{School of Physical Science and Technology, Xinjiang University, Urumqi 830046, China}
\affiliation{Yunnan Observatories, Chinese Academy of Sciences (CAS), Kunming 650216, People’s Republic of China}
\affiliation{Key Laboratory for the Structure and Evolution of Celestial Objects, CAS, Kunming 650216, People’s Republic of China}
\affiliation{International Centre of Supernovae, Yunnan Key Laboratory, Kunming 650216, People’s Republic of China}
\email{guosufen@xju.edu.cn}  

\author{Guoliang Lü*}
\affiliation{School of Physical Science and Technology, Xinjiang University, Urumqi 830046, China}
\affiliation{Xinjiang Astronomical Observatory, Chinese Academy of Sciences, 150 Science 1-Street, Urumqi, Xinjiang 830011, China}
\email{guolianglv@xao.ac.cn} 

\author{Jiao Li}
\affiliation{Yunnan Observatories, Chinese Academy of Sciences (CAS), Kunming 650216, People’s Republic of China}
\affiliation{Key Laboratory for the Structure and Evolution of Celestial Objects, CAS, Kunming 650216, People’s Republic of China}
\affiliation{International Centre of Supernovae, Yunnan Key Laboratory, Kunming 650216, People’s Republic of China}
\email{lijiao@ynao.ac.cn} 

\author{Chunhua Zhu*}
\affiliation{School of Physical Science and Technology, Xinjiang University, Urumqi 830046, China}
\email{zhuchunhua@xju.edu.cn} 

\author{Jianrong Shi}
\affiliation{National Astronomical Observatories, Chinese Academy of Sciences, Beijing, 100101, China}
\email{sjr@bao.ac.cn}
 
\begin{abstract}

Symbiotic stars are among the most crucial binary systems for studying binary star interactions and Type Ia supernova progenitors. Based on the unique observational characteristics of symbiotic stars—strong H\,{\sc I}, He\,{\sc I} emission lines, giant spectral features, and the presence of [O\,{\sc III}], He\,{\sc II}, O\,{\sc VI}, and other emission lines with ionization potentials exceeding 35 eV—and the \textit{Gaia} information, we search for new symbiotic stars using the low-resolution spectroscopic survey data from LAMOST. Thirty-six binary systems have been selected as symbiotic stars or candidates, in which the five known symbiotic stars are included. Among them five systems (ZTF J005917.52+315605.4, ATO J094137.5+075304, LAMOST J200310.90+360822.6, LAMOST J072528.18+342530.4, and V* V758 Cyg) have been found as new symbiotic stars. Notably, LAMOST J072528.18+342530.4 and V* V758 Cyg were also confirmed as new symbiotic stars in a recent study. For the remaining 26 candidates, classification is based solely on the presence of [O\,{\sc III}] emission lines (with ionization potentials $>$ 35 eV) and the absence of He\,{\sc II} high-excitation emission lines. Further observations are needed to confirm their nature as symbiotic stars.

\end{abstract}


\section{INTRODUCTION}\label{INTRODUCTION}

Symbiotic stars were first defined by \citet{merrill1933catalogue} as stellar systems with composite spectra. These spectra feature nebular emission lines, high-energy emission lines, and the continuum spectrum of a late-type giant \citep{merrill1958symbiosis,kenyon1992symbiotic}. The continuous of spectra typically exhibit G-, K-, or M-type giant characteristics, showing prominent molecular absorption bands (e.g., TiO bands in the near-infrared) and atomic absorption lines. The emission lines arise from material ionized by a very hot companion star. Symbiotic stars consist of a hot companion, a cold giant, and a nebula formed by mass transfer from the cold giant to the hot companion via Roche lobe overflow or stellar wind \citep{kenyon1991nature}. The cold giant is usually a red giant, AGB star, or supergiant \citep{murset1999spectral,masetti2006symbiotic,luna2013symbiotic}, while the hot companion is usually a white dwarf \citep{mukai2016lyncis,sokoloski2017flows,akras2019census}. In rare cases, it can be a neutron star \citep{chakrabarty1997symbiotic,masetti2006m,masetti2007x,masetti2011cgcs,lu2012population,bozzo2018igr,yungelson2019wind,merc2019new}, a main-sequence star with an accretion disk \citep{sahai2015pilot}, or a black hole accretor \citep{lopez2017systematic,munari2021galah,lucy2024new}.

\citet{belczynski2000catalogue} put forward an observational definition of symbiotic stars which includes: 1. It has the absorption characteristics of a late-type giant. 2. For most symbiotic stars, there exist strong H\,{\sc I}, He\,{\sc I} emission lines as well as emission spectral lines of ions with an ionization potential of at least 35 eV; for symbiotic stars in outburst, there exist continuous spectra of type A or F and absorption spectral lines from H\,{\sc I}, He\,{\sc I} and single-ionized metals. 3. If there is no late-type giant absorption feature, the emission feature of $\lambda\,6830~$\AA\mbox{ }can also be considered a symbiotic star \citep{kenyon1986symbiotic,mikolajewska1997spectrophotometric}.

Many symbiotic star systems exhibit unique emission features at $\lambda\,6830$\,\AA~and $\lambda\,7088$\,\AA, possibly due to the Raman scattering of O\,{\sc VI} $\lambda\,1032$\,\AA, $\lambda\,1038$\,\AA~resonance lines by neutral hydrogen \citep{nussbaumer1989pharmacological,schmid1989identification}. However, in accretion-only symbiotic stars, where the hot component's luminosity is entirely accretion-powered, the above criteria can not apply. Their spectra are dominated by red giants, with absent optical emission lines but significant UV excess.

The classification of symbiotic stars has evolved significantly since their discovery in 1912. Initially termed "composite spectra" due to unexplained nebular emission lines superimposed on giant continua, these objects were later designated "symbiotic stars" by \citet{merrill1932bright}. \citet{webster1975symbiotic} classified symbiotic stars into S-type (stellar spectra) and D-type (dust masking spectra) based on their infrared characteristic, supplemented by \citet{allen1982infrared}'s D$^\prime$-type featuring mid-infrared excess peaking at 20–30 $\mu$m. \citet{van1993atlas} categorized S/D-types via H$\alpha$ line profiles (e.g., S-1: narrow emissions; S-2: absorption on broad emissions; S-3: strong absorption lines reaching at least the continuum level; D-1: strong, narrow H$\alpha$ emission; D-2: slightly asymmetric H$\alpha$ profiles with enhanced blue emission; D-3: broad emission lines and central absorption features), and \citet{li2015first} defined that the S-type exhibits clear red-giant features with strong Balmer/He emissions but weak forbidden lines; D-type shows obscured giants with diverse forbidden lines; D$^\prime$-type combines D-like emissions with prominent F/G/K giant features. \citet{akras2019census} expanded catalogs and identified infrared-excess S+IR systems.

Theoretically, \citet{skopal2025v1047} simulated the spectral energy distribution of Cen V1047 from near-UV to near-IR and found that Z And-type eruptions can occur not only in symbiotic binaries but also in short-period cataclysmic variables. \citet{tejeda2025geometric} revised the Bondi-Hoyle-Lyttleton (BHL) model by introducing a geometric correction factor, providing a more accurate description of wind accretion in binary star systems. Observationally, \citet{merc2025symbiotic1} has comprehensively reviewed modern ground- and space-based observations of symbiotic stars. This review highlighted the recent increase in the number of symbiotic stars, improvements in classification criteria, and enhanced understanding of their variability. \citet{tatarnikov2025evidence} confirmed 2MASS J21012803+4555377 as a new D-type symbiotic system. \citet{guerrero2025gem} using infrared, optical, ultraviolet, and X-ray data, determined Y Gem is an S-type symbiotic star system. \citet{chen2025new} identified two newly symbiotic stars, i.e., LAMOST J072528.17+342530.4 and V758 Cyg, from the LAMOST DR10 low-resolution spectra. These two stars are also among the new symbiotic stars we have discovered. \citet{stoyanov2024evolution} presented high-resolution spectra of T CrB from 2016--2023, and measured the equivalent widths of H$\alpha$, H$\beta$, He\,{\sc I} and He\,{\sc II} emission lines. \citet{jia2023identifying} who selected 11,709 candidate symbiotic stars from the LAMOST DR9 using machine learning.

Symbiotic star systems provide a natural laboratory for the study of interacting binaries. The hot companion in a symbiotic system accretes material from the cool giant via stellar wind or Roche lobe overflow, offering significant insights into the mass loss from late-type giants, the acceleration mechanisms of stellar winds, and wind accretion \citep{chen2017mass,saladino2019slowly}. The cool companion in symbiotic systems is typically a red giant. Due to its large orbital separation, this systems are crucial for investigating the interactions and evolution between detached and semi-detached binaries \citep{kenyon1986symbiotic,lu2006population}. The hot companion in symbiotic systems is generally a white dwarf with a very high temperature \citep{kenyon1984nature}. Accretion onto the white dwarf can lead to thermonuclear runaway on its surface, resulting in a classical nova outburst \citep{mikolajewska2010symbiotic}. If the white dwarf reaches the Chandrasekhar limit by accreting the material, it can trigger an explosion as a Type Ia supernova \citep{munari1992symbiotic,han2004single,lu2009alternative,di2010progenitors,dilday2012ptf,mikolajewska2011symbiotic,liu2015spectral}. Thus, symbiotic stars provide direct observational samples for studying novae and Type Ia supernovae. Symbiotic stars are also important X-ray sources. Their X-ray emission primarily originates from thermonuclear activity on the hot white dwarf's surface, shock heating from stellar wind collisions, and accretion disk processes.

Consequently, identifying symbiotic stars becomes particularly crucial. \citet{merc2020galactic} compiled and provides an online symbiotic star database in 2020, currently containing 284 symbiotic stars in the Milky Way and 71 in other galaxies, totaling 355. However, this observed number exhibits a stark discrepancy with the theoretical predictions for the symbiotic star population. In 1986, \citet{kenyon1986symbiotic} extrapolated from 150 known symbiotic stars, estimating the population of Galactic symbiotic stars can reach \(4 \times 10^3\). In 1992, \citet{munari1992symbiotic} estimated that the total number of symbiotic stars approximates \(3 \times 10^3\) by assuming symbiotic stars are Type Ia supernova progenitors. One year later, \citet{kenyon1993symbiotic} revised this up to \(3.3 \times 10^3\) by assuming symbiotic stars are candidate progenitors for accretion-driven helium detonation supernovae. In 2002, \citet{magrini2002search} re-estimated the number of Galactic symbiotic stars at \(4 \times 10^3\), based on the assumption that 0.5\% of red giants and AGB stars are in binary systems with white dwarfs. In 2006, \citet{lu2006population} predicted the number of symbiotic stars with accreting white dwarf companions in the Milky Way could range between 1200 and 15,000. The birth rate of symbiotic stars in the Galaxy was estimated at \(0.035\) to \(0.131\) yr\(^{-1}\), implying the total number of symbiotic stars should lie between \(3 \times 10^3\) and \(4 \times 10^4\). Clearly, the actual number of observed symbiotic stars falls significantly less than these of theoretical predictions.

This significant difference might be due to the lack of deep all-sky surveys in terms of wavelengths that effectively distinguish symbiotic stars from other stellar sources \citep{belczynski2000catalogue,miszalski2014identification,mikolajewska2014first}. Most known symbiotic stars have been discovered through large H$\alpha$ emission-line surveys such as the INT Photometric H$\alpha$ Survey (IPHAS) and the AAO/UKST SuperCOSMOS H$\alpha$ Survey (SHS), with subsequent confirmation via deep spectroscopic observations and long-term $I$-band light curve analysis of candidates \citep{corradi2008iphas,corradi2010iphas,rodriguez2014iphas,miszalski2014identification}. The Large Sky Area Multi-Object Fiber Spectroscopic Telescope (LAMOST) survey, being the first spectroscopic survey to release over twenty million spectra, provides exceptional conditions for identifying symbiotic stars with composite spectra. By its Data Release 11 (v1.0) in 2024, LAMOST had released 2.2 times more spectra than the combined total from all other spectroscopic surveys worldwide \citep{luo2012data,luo2015first}. As a result, many peculiar celestial bodies were discovered using LAMOST data, for instance, \citet{yang2025lamost} identified LAMOST J171013.53+2646.0 as a detached, short-period, non-flaring hot subdwarf--white dwarf binary. \citet{kovalev2022spectroscopic} determined radial velocities and stellar parameters using full-spectrum fitting, find a "twin" binary system composed of two nearly identical solar-type stars. \citet{li2015first} confirmed LAMOST J12280490-014825.7 as a symbiotic star.

The main content of this article is the discovery of five new symbiotic stars and 26 symbiotic star candidates from LAMOST. The main structure of the article is as follows: Section \ref{Data} shows the datasets we used. Section \ref{Method} introduces our selection criteria for late-type giants in the datasets, the identification methods of symbiotic systems, and the detailed identification process of symbiotic stars. Section \ref{Result} introduces five newly discovered symbiotic stars and twenty-six new symbiotic star candidates. Section \ref{Conclusion} is our conclusion.

\section{Data}\label{Data}

The Large Sky Area Multi-Object Fiber Spectroscopic Telescope (LAMOST), also known as the Guoshoujing Telescope, is a large optical spectroscopic survey facility independently designed and built in China. It is operated and maintained by the National Astronomical Observatories of the Chinese Academy of Sciences (NAOC). The Guoshoujing Telescope features a unique reflective Schmidt design with 4000 optical fibers positioned at its focal plane, enabling the simultaneous observation of 4000 targets within a 20 square degree field of view, thereby significantly enhancing spectral acquisition efficiency. As of July 2020, LAMOST had completed its Pilot Survey (conducted from October 2011 to June 2012) and the first twelve years of its Regular Survey (commencing September 2012) \citep{zhao2012lamost}. The spectral coverage spans 3700 \AA{} to 9000 \AA{}, matching the observational wavelength range defining symbiotic stars \citep{belczynski2000catalogue}. The spectral resolution at 5500 \AA{} is approximately \(R \approx 1800\) \citep{cui2012large,luo2012data}.

Our research mainly utilized two datasets: 1. The LAMOST Data Release 12 version 1.0 (DR12 v1.0), publicly released on March 26, 2025. This release contains 12,602,390 low-resolution spectra obtained over twelve survey years. These spectra are wavelength and flux calibrated and sky-subtracted, comprising 12,231,890 stellar spectra, 281,059 galaxy spectra, and 89,441 quasar spectra. 2. The subsequent LAMOST Data Release 13 (DR13 v0), released on May 6, 2025. This dataset contains 688,475 low-resolution spectra, also wavelength and flux calibrated and sky-subtracted. Therefore, a total of 13,290,865 low-resolution spectra have been used.

\section{Method}\label{Method}

\begin{figure*}[ht!]
\centering
\includegraphics[width=\textwidth]{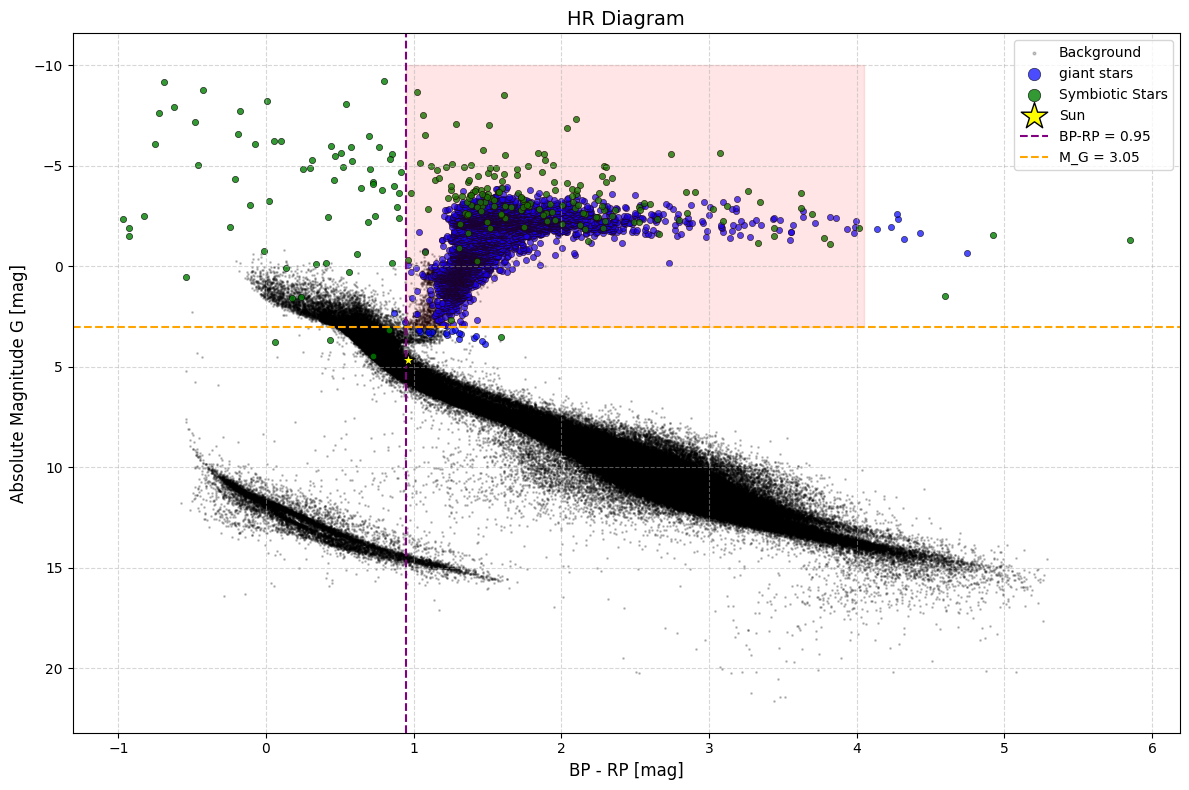}
\caption{Color-magnitude Hertzsprung-Russell diagram. The gray points represent background stars within 125 pc from \textit{\textit{Gaia}}. The green points denote 355 known symbiotic stars \citep{merc2020galactic}. The blue points correspond to 3675 late-type giants \citep{li2022maximum}. The red region indicates the area where symbiotic stars and giants mostly overlap, with $G_{\rm BP} - G_{\rm RP}$ ranging from 0.95 to 4.05 and absolute magnitude $M_{\rm G}$ ranging from -10 to 3.05, which is taken as the screening criteria for late-type giants.}
\label{fig:HR1}
\end{figure*}

\subsection{Selection of late-type giants}\label{Selection Method for Late-Type Giants}

Given that our dataset contains 13,290,865 spectra, exhaustive searches for symbiotic systems would be prohibitive in both time and memory requirements. To optimize identification, we first extract the brighter cool giant components from the full dataset. Our selection methodology for late-type giants from the combined LAMOST DR13 v0.0 and DR12 v1.0 low-resolution spectra is as follows.

We queried \textit{\textit{Gaia}} DR3 for sources within 125\,pc having $\varpi/\sigma_\varpi > 100$ and non-null magnitudes in the $G$, $G_{\text{BP}}$, and $G_{\text{RP}}$ bands, yielding 417,232 objects. Using TOPCAT, we cross-matched the celestial coordinates (RA, DEC) of the 355 symbiotic stars in the current online catalog by \citet{merc2020galactic} with \textit{\textit{Gaia}} DR3 within 1-arcsec. This process yielded \textit{Gaia} DR3 source IDs for 318 counterparts. Query their magnitudes, parallaxes, parallax error, and coordinates data as a background sample of the Hertzsprung-Russell diagram (HRD), facilitating classification of these stars.

\citet{li2022maximum} compiled an intra-galactic giant star catalog using \textit{Gaia} and APASS data, which contains a total of 3675 late-type giants. For these giants, we retrieved $G$, $G_{\rm BP}$, $G_{\rm RP}$ magnitudes, parallaxes, parallax error, and coordinates based on their \textit{Gaia} EDR3 IDs. Both the symbiotic catalog and the late-type giant catalog were extinction corrected using the dustmaps \citep{green20193d} package and plotting on the HRD. Based on their distribution (Figure \ref{fig:HR1}), we defined the selection region for late-type giants.

As shown in Figure \ref{fig:HR1}, based on the significant overlap region between known symbiotics and late-type giants. We set the $G_{\rm BP} - G_{\rm RP}$ color between 0.95 and 4.05, and the absolute $G$-band magnitude ($M_{\rm G}$) between -10 and 3.05 as the selection criteria for late-type giants.

We used the LAMOST Low Resolution Spectral total catalog of DR13 v0 and DR12 v1.0, which includes \textit{Gaia} DR3 source IDs (gaia\_source\_id) from cross-matches performed by LAMOST within 3-arcsec. Using these IDs, we queried \textit{Gaia} DR3 for $G$, $G_{\rm BP}$, $G_{\rm RP}$ magnitudes, parallaxes ($\varpi > 0$), parallax error and coordinates of the sources. After applying extinction correction with the dustmaps, we plotted these sources on the HRD. By defining the selected area ($G_{\rm BP} - G_{\rm RP} \in [0.95, 4.05]$, $M_G \in [-10, 3.05]$), we can derive the final late-type giant sample including 1,061,427 spectra: 16,361 from DR13 v0.0 and 1,045,066 from DR12.

\AtBeginEnvironment{tabular}{\small}
\setlength{\tabcolsep}{3pt}
\renewcommand{\arraystretch}{1.1}

\begin{table*}[ht!]
\centering
\caption{Spectral characteristic lines of LAMOST known symbiotic stars}
\label{tab:spectral_features}
\begin{tabular}{lccc}
\toprule
SIMBAD ID & designation & Emission Lines & Absorption Lines \\
\midrule
EM StHA 190 & LAMOST J214144.88+024354.4 & 
H$\alpha$, H$\beta$, H$\gamma$, H$\delta$, He\,{\sc I}, [O\,{\sc III}], [Ne\,{\sc III}] & 
Ca\,{\sc I}, Ca\,{\sc II}, Fe\,{\sc I}, Na\,{\sc I} \\

EM StHA 169 & LAMOST J194957.58+461520.5 & 
H$\alpha$, H$\beta$, H$\gamma$, H$\delta$, He\,{\sc I}, He\,{\sc II}, [O\,{\sc III}] & 
Ca\,{\sc I}, Ca\,{\sc II}, Fe\,{\sc I}, Na\,{\sc I} \\

UCAC4 441-055195 & LAMOST J122804.90-014825.7 & 
H$\alpha$, H$\beta$, H$\gamma$, H$\delta$, He\,{\sc I}, He\,{\sc II} & 
Ca\,{\sc I}, Ca\,{\sc II}, Fe\,{\sc I}, Na\,{\sc I} \\

IPHAS J184446.08+060703.5 & LAMOST J184446.08+060703.6 & 
H$\alpha$, H$\beta$, H$\gamma$, H$\delta$, He\,{\sc I}, He\,{\sc II}, [O\,{\sc III}] & 
Ca\,{\sc II}, Na\,{\sc I} \\

EM* StHA 32 & LAMOST J043745.63-011911.8 & 
H$\alpha$, H$\beta$, H$\gamma$, H$\delta$, He\,{\sc I}, He\,{\sc II}, [O\,{\sc III}] & 
Ca\,{\sc I}, Ca\,{\sc II}, Fe\,{\sc I}, Na\,{\sc I} \\
\bottomrule
\end{tabular}
{\footnotesize
\tablecomments{This table lists the spectra of five LAMOST symbiotic stars. The first column shows their identifiers in SIMBAD, the second column gives their designation in LAMOST, the third column contains the emission characteristics they exhibit in their spectra, and the fourth column displays the absorption characteristics.}}
\end{table*}

\begin{figure*}[ht!]
\plotone{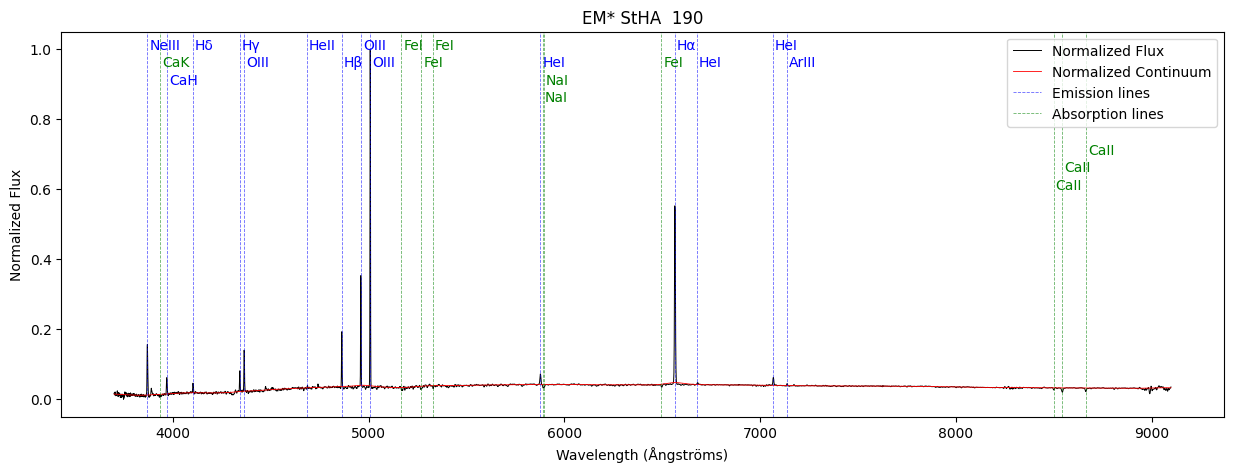}
\caption{The low-resolution LAMOST spectrum of the symbiotic star EM* StHA 190. The black line and the red line represent the spectrum and continuum of the object, respectively.}
\label{fig:method}
\end{figure*}

\subsection{Symbiotic star discrimination criteria}\label{Identification Method for Symbiotic Stars}

For the identification of symbiotic stars using optical spectra, we cross-matched the 355 known symbiotic stars from \citet{merc2020galactic} with the LAMOST data, this process yielded eight confirmed symbiotic stars. After excluding two sources with problem (zero flux at some wavelengths) and one symbiotic star in outburst, we analyzed the remaining five known symbiotic stars. Their common spectral characteristics are shown in Table \ref{tab:spectral_features}, which are consistent with the spectral characteristics and observational definitions outlined in Section \ref{INTRODUCTION}.

Specifically, we adopt the methodology of \citet{li2015first} for the LAMOST data and added the feature lines of symbiotic stars defined by \citet{belczynski2000catalogue} to more effectively identify symbiotic stars. The method is as follows: 1. We normalize the spectra to [0,1] to facilitate batch identification of spectra. 2. We use the laspec \citep{2020ApJS..246....9Z,2021ApJS..256...14Z} package to construct a pseudo-continuum as in Figure \ref{fig:method}. 3. We identify emission lines as regions where the flux exceeds $0.5\sigma$ above the pseudo-continuum (with $\sigma$ being the standard deviation of the pseudo-continuum).

After redshift correction, we measured the following key diagnostic lines:
\begin{itemize}
    \item emission lines: 
    H$\alpha$, H$\beta$, H$\gamma$, H$\delta$, 
    He\,\textsc{i}, O\,\textsc{vi}, He\,\textsc{ii}, 
    [Fe\,\textsc{vii}], [O\,\textsc{iii}], [Ne\,\textsc{iii}]
    
    \item absorption lines: 
    Ca\,\textsc{i}, Ca\,\textsc{ii}, 
    Fe\,\textsc{i}, Na\,\textsc{i}.
\end{itemize}

Finally a late-type giant is classified as symbiotic star if it exhibits strong H\,{\sc I}, He\,{\sc I} emission and at least one high-ionization lines (ionization potential $\geq$35\,eV): [O\,{\sc III}], O\,{\sc VI}, He\,{\sc II}, [Fe\,{\sc VII}], or [Ne\,{\sc III}].

\subsection{Detailed identification process of symbiotic stars}\label{Symbiotic Star Identification Process and Results}

To address this limitation, we leveraged our HRD to screen for a late-type giant sample (Section \ref{Selection Method for Late-Type Giants}). By restricting the analysis to confirmed giants, we ensured the presence of cool companion features while searching for symbiotic-specific composite spectra. This method significantly improved the reliability of symbiotic star identification. Finally, 1,061,427 late-type giants were selected from the combined LAMOST DR13 v0.0 and DR12 v1.0 dataset.

We applied identification criteria described in Section \ref{Identification Method for Symbiotic Stars} to the 1,061,427 giants, requiring detection of strong H\,{\sc I}, He\,{\sc I} emission lines and at least one high-ionization line: [O\,{\sc III}], O\,{\sc VI}, He\,{\sc II}, [Fe\,{\sc VII}], or [Ne\,{\sc III}]. After excluding sources affected by noise and duplicate LAMOST observations of the same target, we obtained 36 symbiotic stars or candidates:
\begin{itemize}
\item Three new symbiotic stars: all exhibit strong H\,{\sc I}, He\,{\sc I} , He\,{\sc II} emission lines, and late-giant absorption features.
\item Two new symbiotic stars that have also been confirmed by \citet{chen2025new} during our analysis period.
\item Five previously known symbiotic stars: symbiotic stars that have been confirmed by \citet{merc2020galactic}.
\item Twenty-six new symbiotic star candidates: all exhibit strong H\,{\sc I}, He\,{\sc I}, [O\,{\sc III}] emission lines, and late-giant absorption features, classified by absorption characteristics:
\begin{itemize}
\item \textit{TiO absorption band:} six new symbiotic star candidates.
\item \textit{Without TiO absorption bands:} twenty new symbiotic star candidates.
\end{itemize}
\end{itemize}
We will discuss our newly confirmed symbiotic stars in two parts. Those that have not been confirmed by others will be discussed in detail in Section \ref{Three newly confirmed symbiotic stars}. Those, recently confirmed by \citet{chen2025new}, will be discussed in detail in Section \ref{Two new symbiotic stars that have recently been confirmed by others}. The newly symbiotic star candidates will be discussed in detail in Section \ref{Symbiotic Star Candidate Analysis}.

\begin{figure*}[ht!]
\plotone{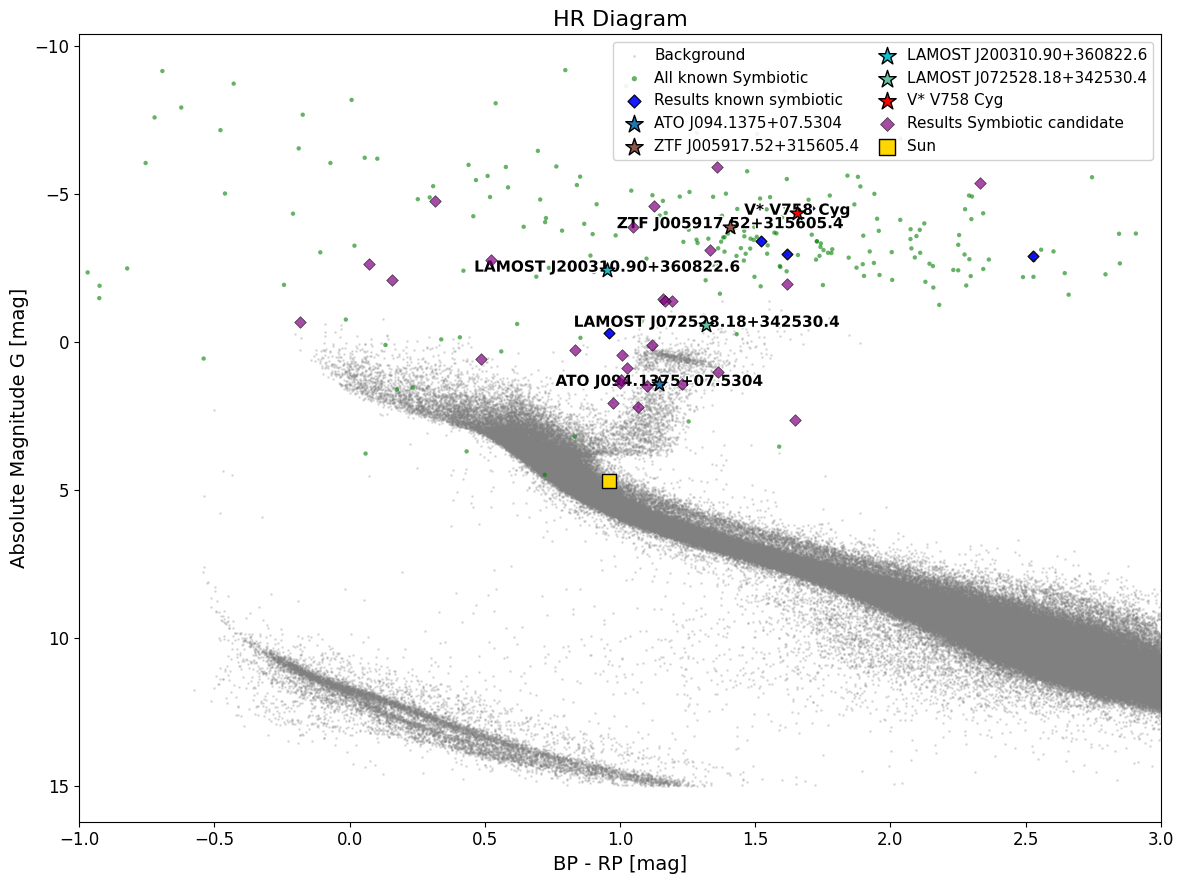}
\caption{Location of our symbiotic star results on the Hertzsprung-Russell diagram. Gray points represent background stars within 125 pc from \textit{Gaia}. Light green points denote the 355 known symbiotic stars from the \citet{merc2020galactic} catalog (hereinafter referred to as the Merc catalog). The yellow rectangle marks the Solar position. Purple diamonds indicate our newly identified candidates. Blue diamonds show known symbiotic stars from the Merc catalog identified in LAMOST. Star symbols represent symbiotic stars identified in this work but not previously listed in the Merc catalog. Notably, during the completion of this work, two star-symbol objects—V* V758 Cyg and LAMOST J072528.18+342530.4—were confirmed as symbiotic stars by \citet{chen2025new}.}
\label{fig:HR2}
\end{figure*}

\section{Results}\label{Result}

This section introduces our newly identified symbiotic stars and symbiotic star candidate stars. The detailed information of these sources are shown in Tables \ref{tab:new_symbiotic_stars}, \ref{tab:known_symbiotic_stars} and \ref{tab:symbiotic_candidate_stars}.

\begin{figure*}[ht!]
\plotone{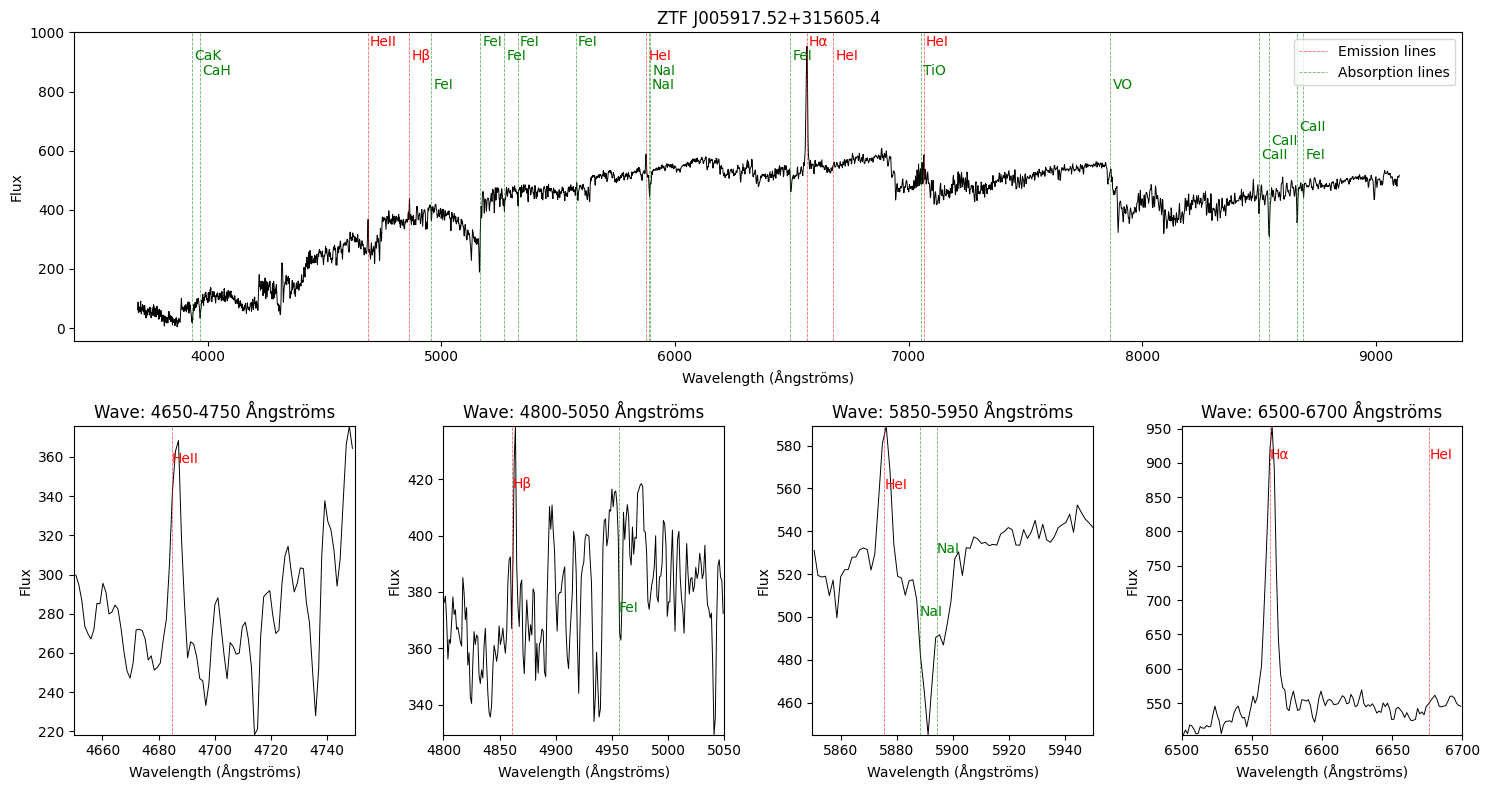}
\caption{The low-resolution LAMOST spectrum of ZTF J005917.52+315605.4. We took four wavelength segments containing the characteristic emission lines of symbiotic stars for amplification. The emission lines are indicated in red, while green for the absorption lines. The details are illustrated in Section \ref{ZTF J005917.52+315605.4}.
}
\label{fig:symbiotic1}
\end{figure*}

\subsection{Three newly confirmed symbiotic stars}\label{Three newly confirmed symbiotic stars}

\subsubsection{ZTF J005917.52+315605.4}\label{ZTF J005917.52+315605.4}

The LAMOST designation of this source is J005917.52+315605.4. The spectrum of ZTF J005917.52+315605.4 is shown in Figure \ref{fig:symbiotic1}. The distinct emission lines, such as, He\,{\sc II} $\lambda$4686\,\AA, H$\beta$ $\lambda$4862\,\AA, He\,{\sc I} $\lambda$5876\,\AA, H$\alpha$ $\lambda$6564\,\AA, He\,{\sc I} $\lambda$6678\,\AA~and He\,{\sc I} $\lambda$7065\,\AA~can be seen, and the emission lines of symbiotic star characteristic are shown with the amplified graph. The TiO absorption band near $\lambda$7054\,\AA~and VO absorption band near $\lambda$7865\,\AA, as well as Ca\,{\sc II K} $\lambda$3934.77\,\AA, Ca\,{\sc II H} $\lambda$3969.59\,\AA, Fe\,{\sc I} $\lambda$4957\,\AA, Fe\,{\sc I} $\lambda$5167.49\,\AA, Fe\,{\sc I} $\lambda$5269.54\,\AA, Fe\,{\sc I} $\lambda$5328.04\,\AA, Na\,{\sc I} $\lambda$5889.91\,\AA, Na\,{\sc I} $\lambda$5895.88\,\AA, Fe\,{\sc I} $\lambda$6494.98\,\AA, Ca\,{\sc II} $\lambda$8500.35\,\AA, Ca\,{\sc II} $\lambda$8544.44\,\AA, Ca\,{\sc II} $\lambda$8664.52\,\AA~can be found.

This source clearly shows the giant continuous spectrum and has multiple molecular absorption bands and absorption lines. \citet{chen2020zwicky} classifies it as a long-period variable, while the LAMSOT pipeline classifies it as carbon type. Its position on the HRD (see Figure \ref{fig:HR2}), along with spectroscopic signatures including TiO absorption bands near $\lambda$7054\,\AA, VO absorption at $\lambda$7865\,\AA~and absorption lines of Fe\,{\sc I}, Na\,{\sc I}, and Ca\,{\sc II}, all of the above information indicate the existence of a late-type giant. The presence of strong H\,{\sc I}, He\,{\sc I}~and high ionization potential He\,{\sc II} (IP $= 54.4\,\text{eV}$) emission lines confirms the existence of a hot stellar component in this source. As its spectrum similar to the symbiotic star EM * StHA 32, we identify it as a new symbiotic stars.

\begin{figure*}[ht!]
\plotone{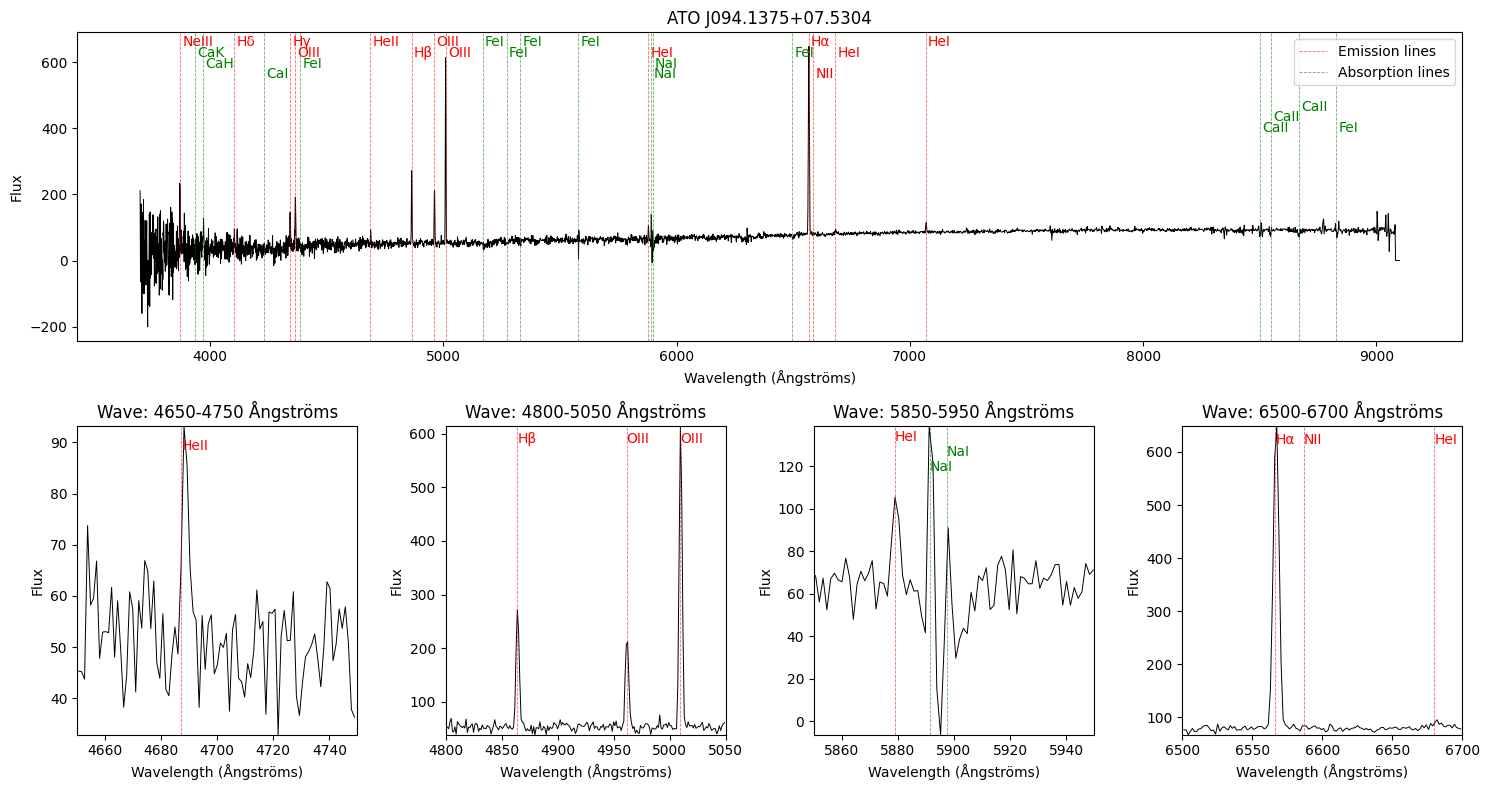}
\caption{Same as Figure \ref{fig:symbiotic1} but for ATO J094.1375+07.5304. The details are discussed in Section \ref{ATO J094.1375+07.5304}.}
\label{fig:symbiotic2}
\end{figure*}

\subsubsection{ATO J094.1375+07.5304}\label{ATO J094.1375+07.5304}

The LAMOST designation of ATO J094.1375+07.5304 is J061633.01+073149.5, and its spectrum is shown in Figure \ref{fig:symbiotic2}. ATO J094.1375+07.5304 displays the emission lines of symbiotic star features: [Ne\,{\sc III}] $\lambda$3869\,\AA, H$\delta$ $\lambda$4102\,\AA, H$\gamma$ $\lambda$4341\,\AA, [O\,{\sc III}] $\lambda$4364\,\AA, He\,{\sc II} $\lambda$4686\,\AA, H$\beta$ $\lambda$4862\,\AA, [O\,{\sc III}] $\lambda$4960\,\AA, [O\,{\sc III}] $\lambda$5008\,\AA, He\,{\sc I} $\lambda$5876\,\AA, H$\alpha$ $\lambda$6564\,\AA, [N\,{\sc II}] $\lambda$6585\,\AA, He\,{\sc I} $\lambda$6678\,\AA, He\,{\sc I} $\lambda$7065\,\AA, as well as Ca\,{\sc II K} $\lambda$3934.77\,\AA, Ca\,{\sc II H} $\lambda$3969.59\,\AA, Ca\,{\sc I} $\lambda$4227.92\,\AA, Fe\,{\sc I} $\lambda$5167.49\,\AA, Fe\,{\sc I} $\lambda$5269.54\,\AA, Fe\,{\sc I} $\lambda$5328.04\,\AA, Na\,{\sc I} $\lambda$5889.91\,\AA, Na\,{\sc I} $\lambda$5895.88\,\AA, Fe\,{\sc I} $\lambda$6494.98\,\AA, Ca\,{\sc II} $\lambda$8500.35\,\AA, Ca\,{\sc II} $\lambda$8544.44\,\AA, Ca\,{\sc II} $\lambda$8664.52\,\AA~absorption lines.

This source exhibits prominent H\,{\sc I}, He\,{\sc I} emission lines, along with He\,{\sc II}, [Ne\,{\sc III}] and [O\,{\sc III}] lines with ionization potentials greater than 35 eV, indicating the presence of a hot component and leading to its identification as a young stellar object (YSO) by \citet{zhang2023catalog1}. However, our study reveals its position on the HRD corresponds to giant stars. Given its classification as a long-period variable star by \citet{heinze2018first} and \citet{chen2020zwicky}, its G5-type designation from the LAMOST pipeline, and the absorption features present in its spectrum, suggesting the existence of a cool companion star. Its spectral resemblance to the symbiotic star EM* StHA 190 further confirms our identification of this system as a new symbiotic star.

\begin{figure*}[ht!]
\plotone{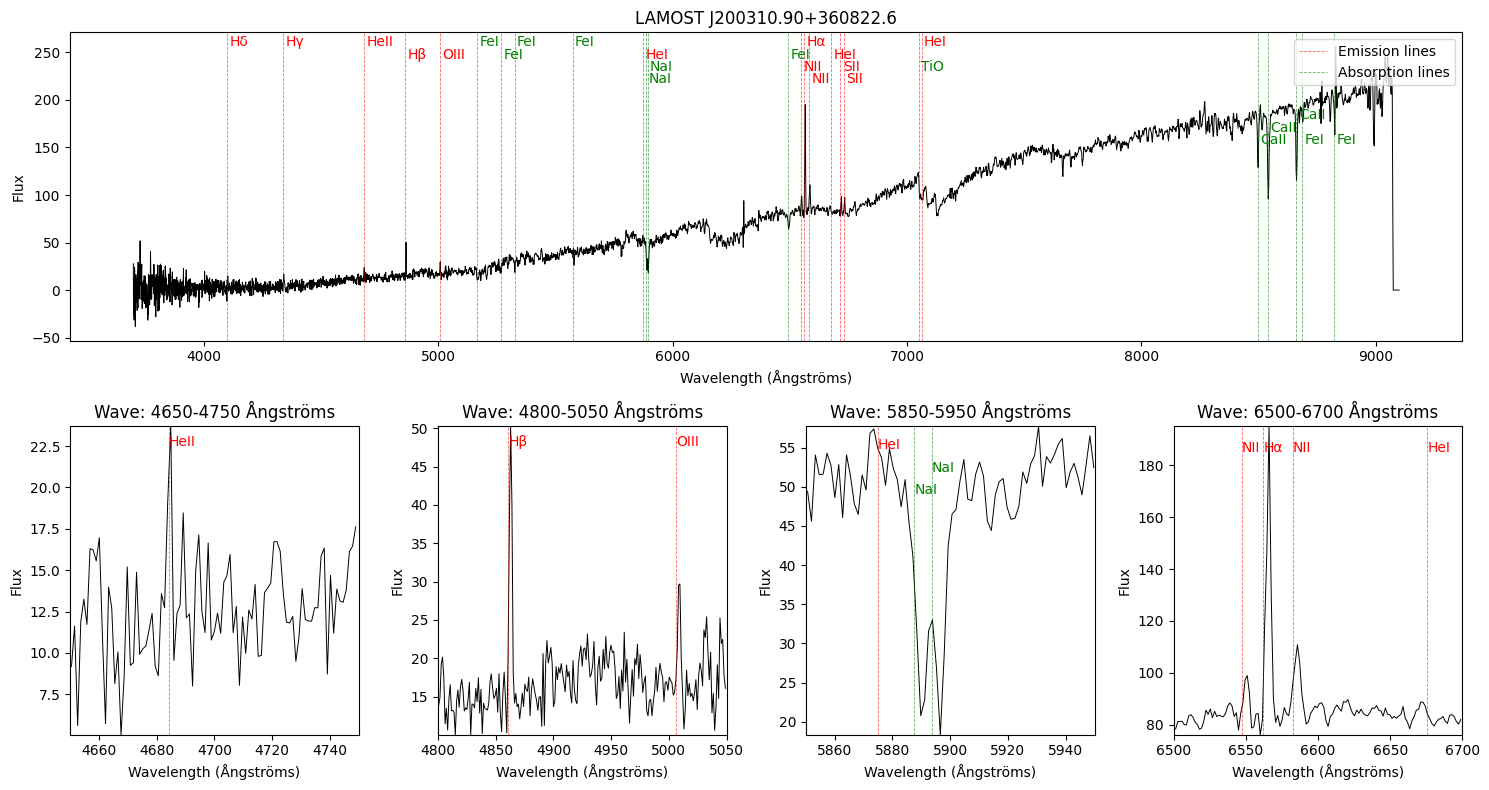}
\caption{Same as Figure \ref{fig:symbiotic1} but for LAMOST J200310.90+360822.6. The details are described in Section \ref{LAMOST J200310.90+360822.6}.}
\label{fig:symbiotic3}
\end{figure*}

\subsubsection{LAMOST J200310.90+360822.6}\label{LAMOST J200310.90+360822.6}

The LAMOST designation of this source is J200310.90 + 360822.6, classified by the LAMOST survey as a stellar source (STAR). Since it was not found in SIMBAD, we temporarily name it LAMOST J200310.90+360822.6. 

The obvious emission lines: H$\delta$ $\lambda$4102\,\AA, H$\gamma$ $\lambda$4341\,\AA, He\,{\sc II} $\lambda$4686\,\AA, H$\beta$ $\lambda$4862\,\AA, [O\,{\sc III}] $\lambda$5008\,\AA, He\,{\sc I} $\lambda$5876\,\AA, [N\,{\sc II}] $\lambda$6549\,\AA, H$\alpha$ $\lambda$6564\,\AA, [N\,{\sc II}] $\lambda$6585\,\AA, He\,{\sc I} $\lambda$6678\,\AA, [S\,{\sc II}] $\lambda$6718\,\AA, [S\,{\sc II}] $\lambda$6732\,\AA, He\,{\sc I} $\lambda$7065\,\AA, and the TiO absorption band near $\lambda$7054\,\AA, as well as absorption lines of Fe\,{\sc I} $\lambda$5167.49\,\AA, Fe\,{\sc I} $\lambda$5269.54\,\AA, Fe\,{\sc I} $\lambda$5328.04\,\AA, Na\,{\sc I} $\lambda$5889.91\,\AA, Na\,{\sc I} $\lambda$5895.88\,\AA, Fe\,{\sc I} $\lambda$6494.98\,\AA, Ca\,{\sc II} $\lambda$8500.35\,\AA, Ca\,{\sc II} $\lambda$8544.44\,\AA, Ca\,{\sc II} $\lambda$8664.52\,\AA, Fe\,{\sc I} $\lambda$8688\,\AA, Fe\,{\sc I} $\lambda$8824\,\AA~can be found as shown in Figure \ref{fig:symbiotic3}.

The significant giant features, including the TiO absorption band and characteristic absorption lines, and the LAMOST pipeline classifies it as gM1 type. Furthermore, based on the aforementioned spectral line characteristics including strong H\,{\sc I} emission, forbidden lines of [N\,{\sc II}] and [S\,{\sc II}], as well as high-ionization emission lines (He\,{\sc II} and [O\,{\sc III}]), so we infer the existence of an additional hot companion star alongside the giant. Therefore, it is classified as a new symbiotic star.

\begin{figure*}[ht!]
\plotone{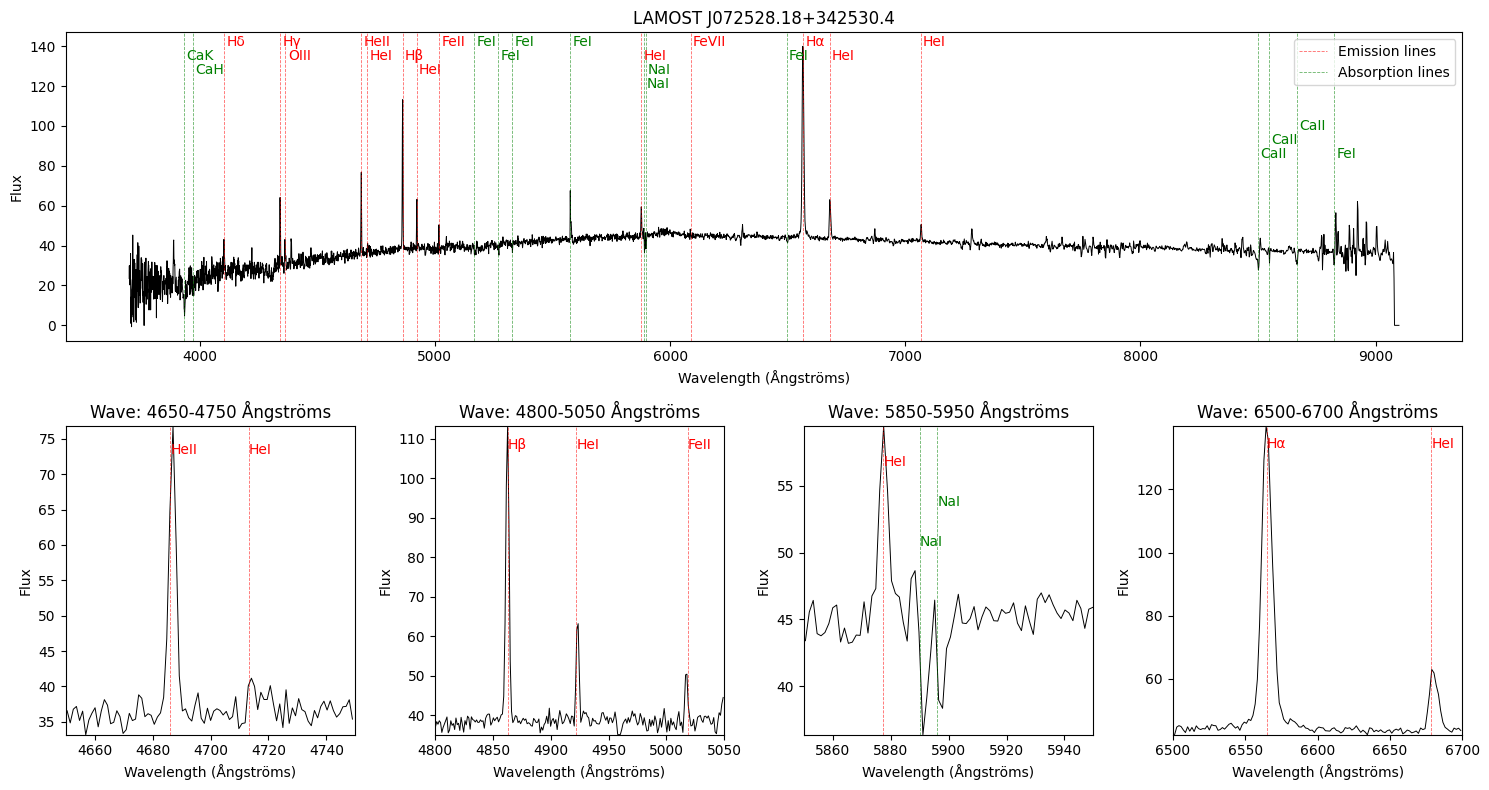}
\caption{Same as Figure 4 but for LAMOST J072528.18+342530.4. The details are presented in Section \ref{LAMOST J072528.18+342530.4}.}
\label{fig:symbiotic4}
\end{figure*}

\subsection{Two recently confirmed symbiotic stars by other researchers}\label{Two new symbiotic stars that have recently been confirmed by others}

\subsubsection{LAMOST J072528.18+342530.4}\label{LAMOST J072528.18+342530.4}

The  LAMOST J072528.17+342530.4 was classification as a K giant by \citet{zhang2023catalog2}, recently, it was identified as a new symbiotic star by \citet{chen2025new} independently. It exhibits strong emission lines including H\,{\sc I}, He\,{\sc I}, He\,{\sc II}, and [O\,{\sc III}] as shown in Figure \ref{fig:symbiotic4}.

\begin{figure*}[ht!]
\plotone{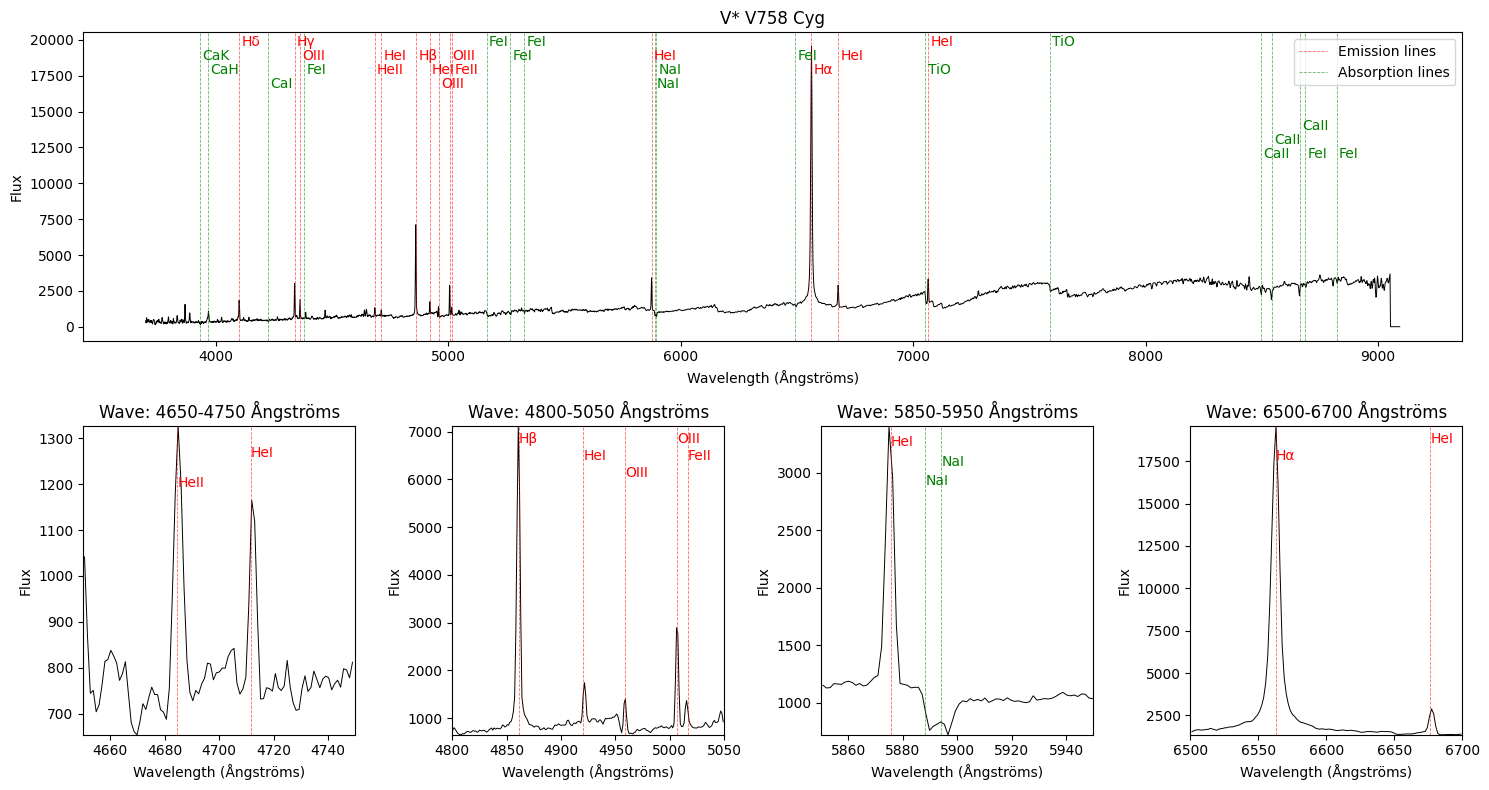}
\caption{Same as Figure 4 but for V* V758 Cyg. The details are shown in Section \ref{V* V758 Cyg}.}
\label{fig:symbiotic5}
\end{figure*}

\subsubsection{V* V758 Cyg}\label{V* V758 Cyg}

The LAMOST designation of V* V758 Cyg is J200023.02+442359.2, and it was classified as a variable star by \citet{pigulski2009all,alfonso2012first,samus2017general,heinze2018first}, a periodic variable by \citet{chen2020zwicky}, and a long-period variable candidate by \citet{lebzelter2023gaia}. This source was simultaneously identified as a symbiotic star by \citet{chen2025new}. It can be seen from Figure \ref{fig:symbiotic5} that the distinct emission lines of H\,{\sc I} , He\,{\sc I}, He\,{\sc II} and [O\,{\sc III}] exist, as well as the TiO absorption bands near $\lambda$7054\,\AA~and $\lambda$7589 \AA.  

Both LAMOST J072528.18+342530.4 526 and V* V758 Cyg are new symbiotic stars independently identified by \citet{chen2025new} and us, which verifies our method.

\subsection{Twenty-six new symbiotic star candidates}\label{Symbiotic Star Candidate Analysis}

In this section, we present and classify 26 symbiotic star candidates identified in this work. These candidates lack He\,{\sc II} $\lambda$4686\,\AA~emission lines, but exhibit strong H\,{\sc I}, He\,{\sc I} and high-ionization [O\,{\sc III}] emission lines, along with late-type giant characteristics. We categorize these candidates into two groups based on the presence of TiO absorption bands: six candidates with prominent giant molecular absorption features and twenty without TiO absorption bands. The detailed information of these sources are presented in Tables \ref{tab:symbiotic_candidate_stars} and \ref{tab:spectral_features_candidates}.

\begin{figure*}[ht!]
\plotone{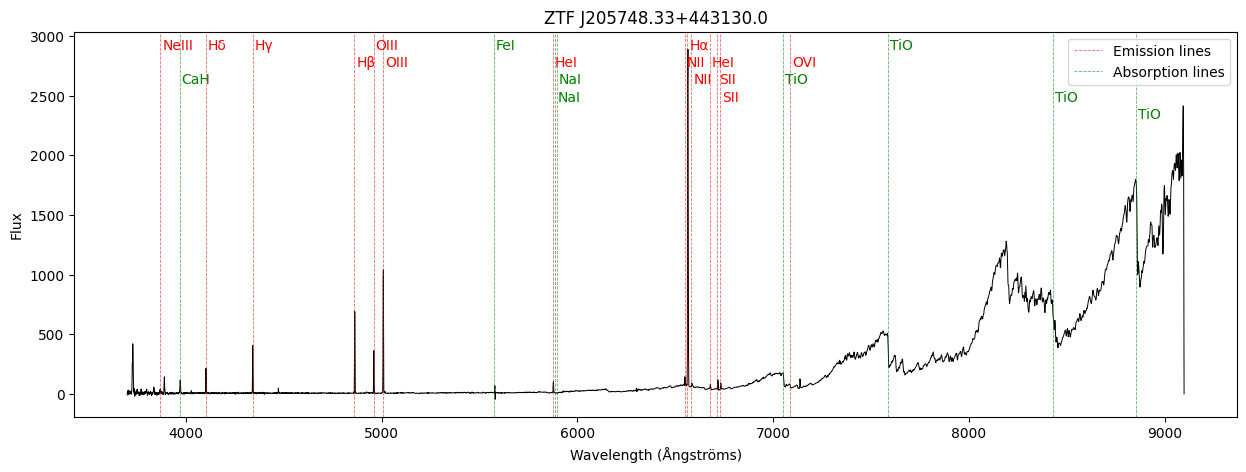}
\caption{Similar to the symbiotic star V347 Nor, the low-resolution LAMOST spectrum of ZTF J205748.33+443130.0 shows comparable features \citep{munari2002multi}. Detailed analysis is presented in Section \ref{Candidates with TiO Absorption Bands}. 
}
\label{fig:symbiotic6}
\end{figure*}

\subsubsection{Candidates with TiO absorption bands}\label{Candidates with TiO Absorption Bands}

Among our symbiotic star candidates, six exhibit prominent TiO absorption characteristics. ZTF J201618.60+375354.5 and ZTF J205748.33+443130.0 have been identified as long-period variables \citep{lebzelter2023gaia,trabucchi2023gaia}, while UCAC4 476-021116 has been classified as an emission-line star \citep{vskoda2020active}, and the remaining three candidates can not be found in the SIMBAD database.

Taking ZTF J205748.33+443130.0 as an example, we plot its spectrum in Figure \ref{fig:symbiotic6}. Its spectrum is very similar to the known symbiotic star V347 Nor \citep[see][Figure 42 on page 49]{munari2002multi}, however, we identify it as a candidate of symbiotic star as there is no obvious He\,{\sc II} $\lambda$4686\,\AA~emission line.

\begin{figure*}[ht!]
\plotone{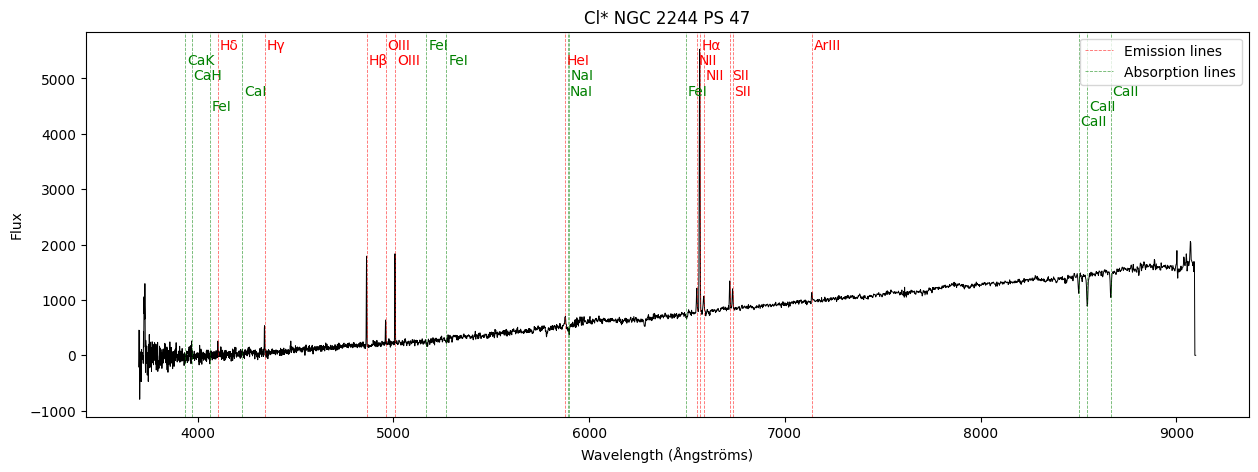}
\caption{Similar to the symbiotic stars KM Vel, V704 Cen, HD 149427, V471 Per, Wray 15-157, AS 201, HD 330036, the low-resolution LAMOST spectrum of Cl* NGC 2244 PS 47 shows comparable features \citep{munari2002multi}. Detailed analysis is shown in Section \ref{Candidates with Metallic Absorption Lines}. }
\label{fig:symbiotic7}
\end{figure*}

\subsubsection{Candidates without TiO absorption bands}\label{Candidates with Metallic Absorption Lines}

There are 20 symbiotic star candidates that only contain metallic absorption lines. Among them, [MJD95] J063133.87+050024.6, 2MASS J05332862-0506019, Cl* NGC 2244 PS 47, and UCAC4 499-030096 have been confirmed as stars \citep{massey1995initial,rebull2000circumstellar,park2002ubvi,kounkel2016spectroscopic}, while the remaining 16 can not be found in the SIMBAD database. The detailed information of these sources are listed in Table \ref{tab:symbiotic_candidate_stars}. 

As an example, we present the spectrum of Cl* NGC 2244 PS 47 in Figure \ref{fig:symbiotic7}. The LAMOST designation of this source is J063134.09+050418.4 and was identified as a star by \textit{Gaia} \citep{park2002ubvi}. Although its spectral characteristics are similar to those of known symbiotic stars \citep[such as V704 Cen, HD 149427, V471 Per, Wray 15-157, AS 201 and KM Vel][]{munari2002multi}, i.e., strong H\,{\sc I}, He\,{\sc I} and [O\,{\sc III}] emission lines, we class it as a candidate of symbiotic star due to lack of He\,{\sc II} $\lambda$4686\,\AA~emission line.

\section{Conclusion}\label{Conclusion}

According to the method of selecting symbiotic stars from LAMOST proposed by \citet{li2015first}, we not only identify H$\alpha$ and H$\beta$ lines, but also add the characteristic emission/absorption lines observed in symbiotic stars: emission lines include H$\gamma$, H$\delta$, He\,{\sc I}, O\,{\sc VI}, He\,{\sc II}, [Fe\,{\sc VII}], [O\,{\sc III}], [Ne\,{\sc III}]; absorption lines include Ca\,{\sc I}, Fe\,{\sc I}, Na\,{\sc I}, Ca\,{\sc II}. Based on the broad definition of symbiotic stars \citep{belczynski2000catalogue}, we set screening criteria specifically for symbiotic stars. Applying our method to 13,302,574 LAMOST low-resolution spectra (from DR13v0, DR12v1.0, and machine learning symbiotic star candidates) \citep{jia2023identifying}, we first use the HRD to select 1,061,427 late-type giant spectra for feature line intensity identification and symbiotic star screening. Finally, we obtain 36 symbiotic stars or candidates:

\begin{itemize}
    \item Three new symbiotic stars: ZTF J005917.52+315605.4, ATO J094.1375+07.5304, and LAMOST J200310.90+360822.6 (designated temporarily as they are not listed in the SIMBAD database).
    \item Two recently confirmed symbiotic stars: LAMOST J072528.18+342530.4, V758 Cyg \citep{chen2025new}.
    \item Twenty-six symbiotic star candidates.
    \item Five known symbiotic stars: EM StHA 190, UCAC4 44-055195, HD 342007, EM StHA 169, EM StHA 32.
\end{itemize}

The recent confirmation of two symbiotic stars by \citet{chen2025new} matches two of the five new symbiotic stars identified in this work. Combined with the five known symbiotic stars we identified, these consistencies further verify the reliability of our method. The observed number of discovered symbiotic stars significantly deviates from the theoretical predictions, suggesting many undetected symbiotic star exist. Among the 13,302,574 LAMOST spectra analyzed, we identified five new symbiotic stars within our restricted late-type giant selection criteria. As shown in Figure \ref{fig:HR1}, numerous symbiotic stars likely reside outside our current selection range. Future work should expand the giant star selection criteria to encompass all known symbiotic stars, then apply our method to systematically search for new symbiotic stars.

\begin{acknowledgments}
This work received the support of the Nation Natural Science Foundation of China under grants 12463011, 12003025, 12163005, 12373038, and 12288102; the Natural Science Foundation of Xinjiang No. 2024D01C230, No. 2022TSYCLJ0006 and 2022D01D85;  the China Manned Space Program with grant No. CMS-CSST-2025-A15. Guoshoujing Telescope (the Large Sky Area Multi-Object Fiber Spectroscopic Telescope, LAMOST) is a National Major Scientific Project built by the Chinese Academy of Sciences. Funding for the Project has been provided by the National Development and Reform Commission. LAMOST is operated and managed by the National Astronomical Observatories, Chinese Academy of Sciences. This research makes use of data from the European Space Agency (ESA) mission \textit{Gaia} (\url{https://www.cosmos.esa.int/gaia}), processed by the \textit{Gaia} Data Processing and Analysis Consortium (DPAC; \url{https://www.cosmos.esa.int/web/gaia/dpac/consortium}). Funding for the DPAC has been provided by national institutions, in particular the institutions participating in the \textit{Gaia} Multilateral Agreement. This research also makes use of the SIMBAD database \citep{wenger2000simbad}, operated at Centre de Donnees astronomiques de Strasbourg (CDS), France. This publication made also use of many software packages in Python.

\textit{Software}: Astropy \citep{robitaille2013astropy}, TOPCAT \citep{shopbell2005astronomical}, dustmaps \citep{green20193d}, laspec \citep{2020ApJS..246....9Z,2021ApJS..256...14Z}, Matplotlib \citep{Hunter:2007}, NumPy \citep{harris2020array}, SciPy \citep{2020SciPy-NMeth}.
\end{acknowledgments}

\bibliography{TEXT1}{}
\bibliographystyle{aasjournalv7}

\appendix

\section{The fiber mask problem in LAMOST data}\label{Data fiber problem}
\begin{figure*}[ht!]
\plotone{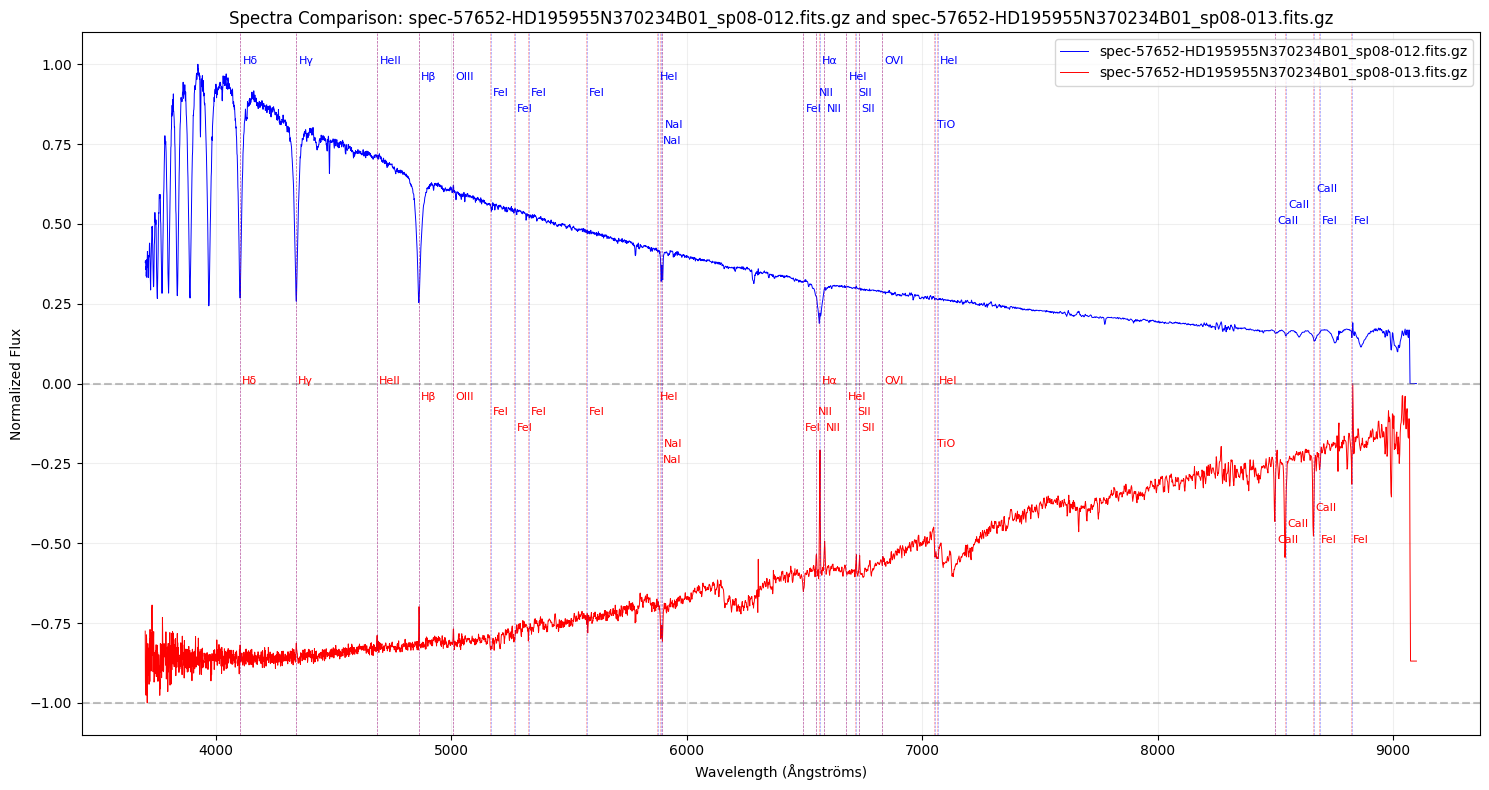}
\caption{Spectral comparison diagram of adjacent fibers LAMOST J072439.08+341202.7 fiberid 193 and LAMOST J072528.17+342530.4 fiberid 194. Spectral comparison diagram of adjacent fibers LAMOST J200321.88+361111.2 fiberid 12 and LAMOST J200310.90+360822.6 fiberid 13. There are no identical or identical parts between the two spectra, so the spectrum of our source has not been contaminated by the adjacent optical fiber. For a detailed description, please refer to Section \ref{LAMOST J200310.90+360822.6}.}
\label{fig:symbiotic3_Mask}
\end{figure*}

The observation parameter keyword "FIB\_MASK" in the FITS file of LAMOST J072439.08+341202.7 is 128, indicating a possible issue with the optical fiber. Convert the decimal "FIB\_MASK=128" to a nine-bit binary number, which is 10000000. The 8th position indicates the existence of a NEARWHOPPER problem, meaning that this is a fiber with a huge adjacent traffic volume. Therefore, we found that the spid of this source FITS file is 5 and the fiberid is 194. After checking the FITS data adjacent to this fiber observed this time, we indeed only found that FIB\_MASK=64 exists in the FITS file with fiberid 193, which is converted to a nine-bit binary number of 1,000,000. The 7th position indicates the existence of the WHOPPER problem, meaning that this optical fiber has a flow rate more than 15 times that of the adjacent optical fiber (a fiber with a huge flow rate). Subsequently, we compared the spectra of these two problem optical fibers, as shown in Figure \ref{fig:symbiotic3_Mask}. Eventually, it was found that the spectrum of our source was not contaminated by the spectra of the adjacent optical fibers.

\begin{figure*}[ht!]
\plotone{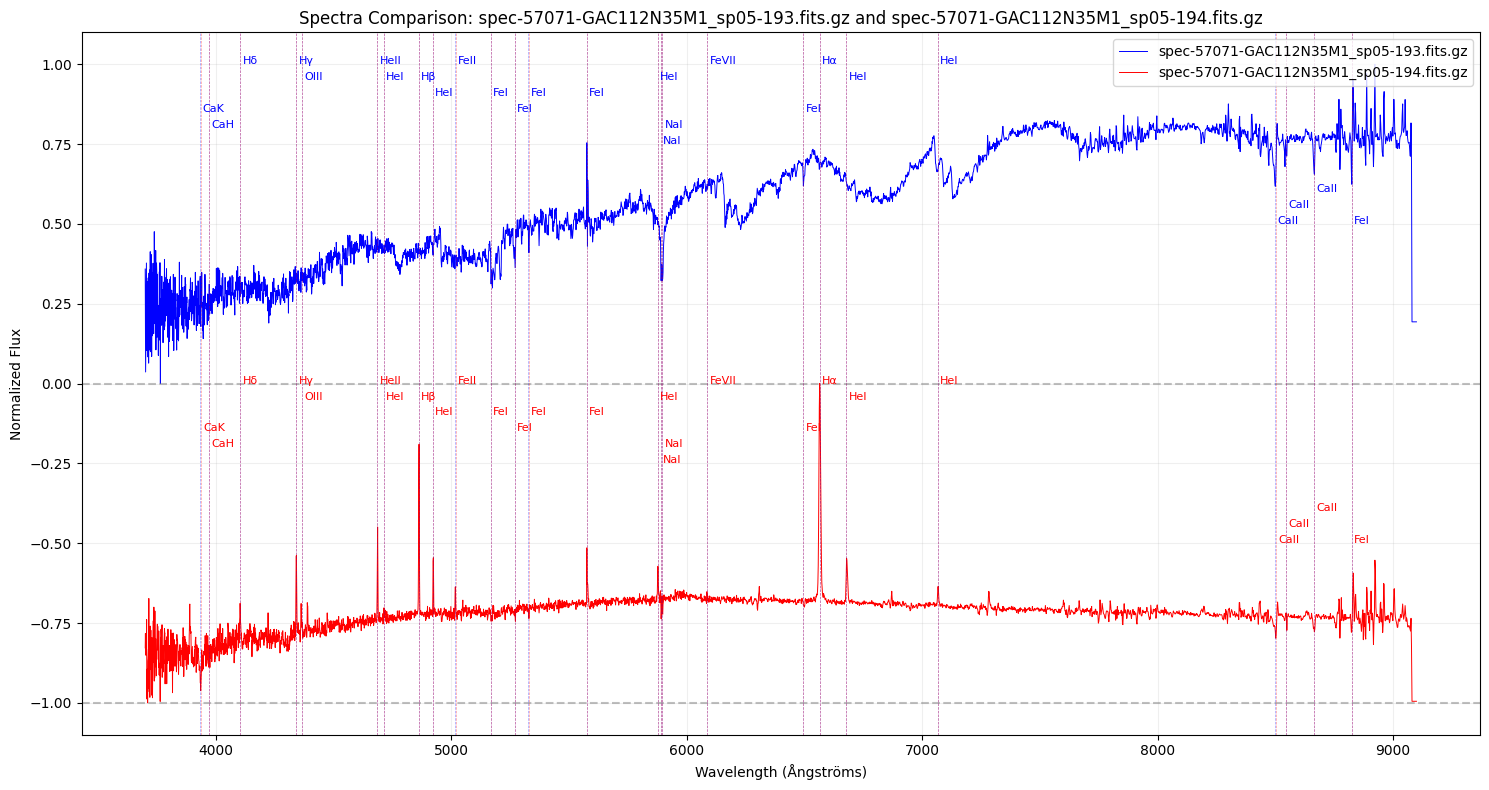}
\caption{Spectral comparison diagram of adjacent fibers LAMOST J200321.88+361111.2 fiberid 12 and LAMOST J200310.90+360822.6 fiberid 13. There are no identical or identical parts between the two spectra, so the spectrum of our source has not been contaminated by the adjacent optical fiber. For a detailed description, please refer to Section \ref{LAMOST J072528.18+342530.4}.}
\label{fig:symbiotic4_Mask}
\end{figure*}

The observation parameter keyword "FIB\_MASK" in the FITS file of LAMOST J200310.90+360822.6 is 128, the same as the previous source. Therefore, we found that the spid of this source FITS file is 8 and the fiberid is 13. After checking the FITS data adjacent to this fiber observed this time, we indeed only found that FIB\_MASK=64 exists in the FITS file with fiberid 12, the same as the previous source. Subsequently, we compared the spectra of these two problem optical fibers, as shown in Figure \ref{fig:symbiotic4_Mask}. Eventually, it was found that the spectrum of our source was not contaminated by the spectra of the adjacent optical fibers.

\section{Catalog of symbiotic stars and candidates}\label{Catalog of symbiotic stars and candidates}
\begin{table*}[ht!]
\centering
\tabletypesize{\scriptsize}
\caption{Five new symbiotic stars}
\begin{tabular}{l r r r r r r r r}
\hline
\colhead{LAMOST} & \colhead{RA} & \colhead{Dec} & \colhead{$g$} & \colhead{$G_{\rm BP}$} & \colhead{$G_{\rm RP}$} & \colhead{RV} & \colhead{$T_{\rm eff}$} & \colhead{$\log g$} \\
\colhead{designation} & \colhead{(deg)} & \colhead{(deg)} & \colhead{(mag)} & \colhead{(mag)} & \colhead{(mag)} & \colhead{(km s$^{-1}$)} & \colhead{(K)} & \colhead{(dex)} \\
\hline
J005917.52+315605.4 & 14.8230161 & 31.9348381 & 14.385245 & 15.086717 & 13.585769 & -64.15396 & 4396.8584 & 1.7188 \\
J061633.01+073149.5 & 94.137551 & 7.5304271 & 15.944861 & 16.945942 & 14.9922695 & - & 4970.015 & 3.6583 \\
J200310.90+360822.6 & 300.795455 & 36.139635 & 15.543657 & 17.426613 & 14.259249 & -123.68271 & 3710.99 & 0.986 \\
J072528.17+342530.4 & 111.36739 & 34.425122 & 17.296923 & 17.937033 & 16.52981 & - & 4390.8687 & 4.3557 \\
J200023.02+442359.2 & 300.095917 & 44.399778 & 12.619182 & 14.073226 & 11.429489 & -87.81241 & 3661.5657 & 0.7918 \\
\hline
\end{tabular}
\label{tab:new_symbiotic_stars}
{\footnotesize
\tablecomments{The catalogue of five new symbiotic stars. \texttt{LAMOST designation} is LAMOST unique source designation, \texttt{RA} is right ascension, \texttt{Dec} is declination, \texttt{$G$} is \textit{Gaia} $G$-band mean magnitude, \texttt{$G_{\rm BP}$} is \textit{Gaia} integrated BP mean magnitude, \texttt{$G_{\rm RP}$} is \textit{Gaia} integrated RP mean magnitude, \texttt{RV} is \textit{Gaia}'s radial velocity, \texttt{$T_{\rm eff}$} is \textit{Gaia} effective temperature (from GSP-Phot Aeneas best library using BP/RP spectra) and LAMOST effective temperature (obtained by the LASP), \texttt{$\log g$} is \textit{Gaia} surface gravity (from GSP-Phot Aeneas best library using BP/RP spectra) and LAMOST Surface gravity (obtained by the LASP).}}
\end{table*}

\begin{table*}[ht!]
\centering
\tabletypesize{\scriptsize}
\caption{Five known symbiotic stars}
\begin{tabular}{l r r r r r r r r}
\hline
\colhead{LAMOST} & \colhead{RA} & \colhead{Dec} & \colhead{$g$} & \colhead{$G_{\rm BP}$} & \colhead{$G_{\rm RP}$} & \colhead{RV} & \colhead{$T_{\rm eff}$} & \colhead{$\log g$} \\
\colhead{designation} & \colhead{(deg)} & \colhead{(deg)} & \colhead{(mag)} & \colhead{(mag)} & \colhead{(mag)} & \colhead{(km s$^{-1}$)} & \colhead{(K)} & \colhead{(dex)} \\
\hline
J194957.59+461520.5 & 297.48996 & 46.255718 & 12.488104 & 13.44773 & 11.4545965 & - & 3670.73 & 0.775 \\
J214144.88+024354.4 & 325.437011 & 2.73178 & 10.262842 & 10.731504 & 9.600878 & -3.1831923 & 5129.56 & 2.099 \\
J043745.63-011911.8 & 69.44013 & -1.319971 & 12.286708 & 13.029832 & 11.460179 & 326.4626 & 4399.6714 & 1.3572 \\
J122804.90-014825.7 & 187.0204458 & -1.8071583 & 12.045676 & 12.839189 & 11.176297 & 368.99344 & 4148.83 & 1.206 \\
J182207.84+232719.9 & 275.532704 & 23.455544 & 9.933405 & 11.455687 & 8.67594 & - & 3408.95 & 0.461 \\
\hline
\end{tabular}
\label{tab:known_symbiotic_stars}
{\footnotesize 
\tablecomments{The catalog of five known symbiotic stars, its contents shown are the same as those in Table \ref{tab:new_symbiotic_stars}.}}
\end{table*}

\begin{table*}[ht!]
\centering
\tabletypesize{\scriptsize}
\caption{Twenty-six new symbiont star candidates}
\begin{tabular}{l r r r r r r r r}
\hline
\colhead{LAMOST} & \colhead{RA} & \colhead{Dec} & \colhead{$g$} & \colhead{$G_{\rm BP}$} & \colhead{$G_{\rm RP}$} & \colhead{RV} & \colhead{$T_{\rm eff}$} & \colhead{$\log g$} \\
\colhead{designation} & \colhead{(deg)} & \colhead{(deg)} & \colhead{(mag)} & \colhead{(mag)} & \colhead{(mag)} & \colhead{(km s$^{-1}$)} & \colhead{(K)} & \colhead{(dex)} \\
\hline
J063148.01+051037.6 & 97.950044 & 5.177135 & 14.437584 & 16.723442 & 13.069414 & 104.420204 & 3505.39 & 0.461 \\
J063134.09+050418.4 & 97.892045 & 5.071799 & 14.589907 & 16.02964 & 13.429913 & 40.64758 & - & - \\
J040533.87+354741.6 & 61.391158 & 35.794907 & 15.504053 & 16.362303 & 14.5909 & - & 4735.358 & 3.0895 \\
J040409.93+354156.3 & 61.041379 & 35.698993 & 15.641661 & 16.366732 & 14.80805 & - & 4793.01 & 2.805 \\
J035820.30+361009.1 & 59.584599 & 36.1692 & 15.785639 & 16.564688 & 14.918266 & - & 4934.041 & 3.6019 \\
J040252.69+355132.3 & 60.719553 & 35.858994 & 15.725463 & 16.450855 & 14.89326 & - & 5289.675 & 3.5708 \\
J201618.58+375354.4 & 304.0774293 & 37.8984669 & 13.824789 & 17.279465 & 12.302737 & -21.710274 & 3333.94 & 0.372 \\
J052506.90+334102.6 & 81.278768 & 33.684083 & 15.788467 & 16.863514 & 14.767342 & - & 6288.883 & 3.3218 \\
J035934.45+352505.9 & 59.893576 & 35.418311 & 16.887276 & 17.467104 & 16.163868 & - & 4858.2812 & 4.3619 \\
J040443.64+353847.0 & 61.181861 & 35.646404 & 16.891907 & 17.478514 & 15.965531 & - & 4687.1353 & 4.3856 \\
J202350.36+403845.4 & 305.959869 & 40.645945 & 15.756848 & 18.170433 & 14.35553 & -5.5397987 & - & - \\
J202507.89+430609.7 & 306.282898 & 43.102712 & 15.518458 & 18.177391 & 14.0839 & -8.766446 & 3711.32 & 0.776 \\
J053801.58+101745.6 & 84.506608 & 10.296019 & 14.830967 & 15.625402 & 13.957741 & 36.722633 & 4845.5415 & 2.7243 \\
J053709.99+095729.9 & 84.29164 & 9.9583232 & 15.608242 & 16.566416 & 14.641755 & - & 4989.3647 & 3.2275 \\
J053328.62-050601.9 & 83.369275 & -5.1005419 & 15.144981 & 16.463715 & 14.021867 & 32.579758 & 5058.7437 & 3.2392 \\
J205832.75+442428.4 & 314.636473 & 44.407902 & 14.395758 & 16.47094 & 13.067833 & -77.95555 & 3740.04 & 1.184 \\
J205748.33+443130.0 & 314.451379 & 44.525017 & 15.292596 & 19.770132 & 13.635615 & - & 3604.2837 & 0.8953 \\
J063905.86+094650.8 & 99.774435 & 9.780805 & 17.018427 & 18.41442 & 15.885577 & - & 4970.321 & 3.7827 \\
J063922.33+094321.5 & 99.843068 & 9.722665 & 16.72781 & 18.229435 & 15.553658 & - & 5186.4473 & 3.6242 \\
J064153.35+103842.6 & 100.47233 & 10.645186 & 17.124338 & 18.363811 & 16.034163 & - & - & - \\
J195420.64+384158.9 & 298.58604 & 38.699702 & 14.924134 & 15.829988 & 13.991077 & -26.63061 & - & - \\
J063257.79+051427.2 & 98.2408318 & 5.240911 & 14.977341 & 16.612865 & 13.751184 & 63.106224 & 15005.335 & 3.6496 \\
J063133.88+050024.3 & 97.8911706 & 5.0067575 & 15.587823 & 16.124918 & 14.862849 & - & 9912.657 & 3.8178 \\
J035745.75+362336.2 & 59.440628 & 36.3934013 & 16.478077 & 16.982618 & 15.807557 & - & 5696.4 & 3.938 \\
J053008.24+095043.9 & 82.5343716 & 9.8455385 & 17.196892 & 17.83023 & 16.067722 & - & 4317.03 & 4.741 \\
J063827.70+102829.1 & 99.6154246 & 10.4747737 & 14.942393 & 16.010424 & 13.922214 & 66.34043 & 6044.739 & 3.0685 \\
\hline
\end{tabular}
\label{tab:symbiotic_candidate_stars}
{\footnotesize
\tablecomments{The catalogue of twenty-six new symbiotic star candidates, its contents shown are the same as those in Table \ref{tab:new_symbiotic_stars}.}}
\end{table*}

\begin{table*}[ht!]
\centering
\caption{Spectral characteristic lines of symbiotic star candidates}
\label{tab:spectral_features_candidates}
\begin{tabular}{lccc}
\toprule
Designation & Emission Lines & Molecular absorption band & Absorption Lines \\
\midrule
J063148.01+051037.6 & 
H$\alpha$, H$\beta$, H$\gamma$, H$\delta$, He\,{\sc I}, [O\,{\sc III}] & 
TiO & 
Ca\,{\sc I}, Fe\,{\sc I}, Na\,{\sc I}, Ca\,{\sc II} \\

J063134.09+050418.4 & 
H$\alpha$, H$\beta$, H$\gamma$, H$\delta$, He\,{\sc I}, [O\,{\sc III}] & 
-- & 
Ca\,{\sc I}, Fe\,{\sc I}, Na\,{\sc I}, Ca\,{\sc II} \\

J040533.87+354741.6 & 
H$\alpha$, H$\beta$, H$\gamma$, H$\delta$, He\,{\sc I}, [O\,{\sc III}] & 
-- & 
Ca\,{\sc I}, Fe\,{\sc I}, Na\,{\sc I}, Ca\,{\sc II} \\

J040409.93+354156.3 & 
H$\alpha$, H$\beta$, H$\gamma$, H$\delta$, He\,{\sc I}, [O\,{\sc III}] & 
-- & 
Ca\,{\sc I}, Fe\,{\sc I}, Na\,{\sc I}, Ca\,{\sc II} \\

J035820.30+361009.1 & 
H$\alpha$, H$\beta$, H$\gamma$, H$\delta$, He\,{\sc I}, [O\,{\sc III}] & 
-- & 
Ca\,{\sc I}, Fe\,{\sc I}, Na\,{\sc I}, Ca\,{\sc II} \\

J040252.69+355132.3 & 
H$\alpha$, H$\beta$, H$\gamma$, H$\delta$, He\,{\sc I}, [O\,{\sc III}] & 
-- & 
Ca\,{\sc I}, Fe\,{\sc I}, Na\,{\sc I}, Ca\,{\sc II} \\

J201618.58+375354.4 & 
H$\alpha$, H$\beta$, H$\gamma$, H$\delta$, He\,{\sc I}, [O\,{\sc III}] & 
TiO & 
Ca\,{\sc I}, Fe\,{\sc I}, Na\,{\sc I}, Ca\,{\sc II} \\

J052506.90+334102.6 & 
H$\alpha$, H$\beta$, H$\gamma$, H$\delta$, He\,{\sc I}, [O\,{\sc III}] & 
-- & 
Ca\,{\sc I}, Fe\,{\sc I}, Na\,{\sc I}, Ca\,{\sc II} \\

J035934.45+352505.9 & 
H$\alpha$, H$\beta$, H$\gamma$, H$\delta$, He\,{\sc I}, [O\,{\sc III}] & 
-- & 
Ca\,{\sc I}, Fe\,{\sc I}, Na\,{\sc I}, Ca\,{\sc II} \\

J040443.64+353847.0 & 
H$\alpha$, H$\beta$, H$\gamma$, H$\delta$, He\,{\sc I}, [O\,{\sc III}] & 
-- & 
Ca\,{\sc I}, Fe\,{\sc I}, Na\,{\sc I}, Ca\,{\sc II} \\

J202350.36+403845.4 & 
H$\alpha$, H$\beta$, H$\gamma$, H$\delta$, He\,{\sc I}, [O\,{\sc III}] & 
TiO, VO & 
Ca\,{\sc I}, Fe\,{\sc I}, Na\,{\sc I}, Ca\,{\sc II} \\

J202507.89+430609.7 & 
H$\alpha$, H$\beta$, H$\gamma$, H$\delta$, He\,{\sc I}, [O\,{\sc III}] & 
TiO & 
Ca\,{\sc I}, Fe\,{\sc I}, Na\,{\sc I}, Ca\,{\sc II} \\

J053801.58+101745.6 & 
H$\alpha$, H$\beta$, H$\gamma$, H$\delta$, He\,{\sc I}, [O\,{\sc III}] & 
-- & 
Ca\,{\sc I}, Fe\,{\sc I}, Na\,{\sc I}, Ca\,{\sc II} \\

J053709.99+095729.9 & 
H$\alpha$, H$\beta$, H$\gamma$, H$\delta$, He\,{\sc I}, [O\,{\sc III}] & 
-- & 
Ca\,{\sc I}, Fe\,{\sc I}, Na\,{\sc I}, Ca\,{\sc II} \\

J053328.62-050601.9 & 
H$\alpha$, H$\beta$, H$\gamma$, H$\delta$, He\,{\sc I}, [O\,{\sc III}] & 
-- & 
Ca\,{\sc I}, Fe\,{\sc I}, Na\,{\sc I}, Ca\,{\sc II} \\

J205832.75+442428.4 & 
H$\alpha$, H$\beta$, H$\gamma$, H$\delta$, He\,{\sc I}, [O\,{\sc III}] & 
TiO & 
Fe\,{\sc I}, Na\,{\sc I}, Ca\,{\sc II} \\

J205748.33+443130.0 & 
H$\alpha$, H$\beta$, H$\gamma$, H$\delta$, He\,{\sc I}, [O\,{\sc III}], [Ne\,{\sc III}] & 
TiO & 
Fe\,{\sc I}, Na\,{\sc I}, Ca\,{\sc II} \\

J063905.86+094650.8 & 
H$\alpha$, H$\beta$, H$\gamma$, H$\delta$, He\,{\sc I}, [O\,{\sc III}] & 
-- & 
Ca\,{\sc I}, Fe\,{\sc I}, Na\,{\sc I}, Ca\,{\sc II} \\

J063922.33+094321.5 & 
H$\alpha$, H$\beta$, H$\gamma$, H$\delta$, He\,{\sc I}, [O\,{\sc III}] & 
-- & 
Fe\,{\sc I}, Na\,{\sc I}, Ca\,{\sc II} \\

J064153.35+103842.6 & 
H$\alpha$, H$\beta$, H$\gamma$, H$\delta$, He\,{\sc I}, [O\,{\sc III}] & 
-- & 
Ca\,{\sc I}, Fe\,{\sc I}, Na\,{\sc I}, Ca\,{\sc II} \\

J195420.64+384158.9 & 
H$\alpha$, H$\beta$, H$\gamma$, H$\delta$, He\,{\sc I}, [O\,{\sc III}] & 
-- & 
Ca\,{\sc I}, Fe\,{\sc I}, Na\,{\sc I}, Ca\,{\sc II} \\

J063257.79+051427.2 & 
H$\alpha$, H$\beta$, H$\gamma$, H$\delta$, He\,{\sc I}, [O\,{\sc III}] & 
-- & 
Ca\,{\sc I}, Fe\,{\sc I}, Na\,{\sc I}, Ca\,{\sc II} \\

J063133.88+050024.3 & 
H$\alpha$, H$\beta$, H$\gamma$, H$\delta$, He\,{\sc I}, [O\,{\sc III}], [Ne\,{\sc III}] & 
-- & 
Ca\,{\sc I}, Fe\,{\sc I}, Na\,{\sc I}, Ca\,{\sc II} \\

J035745.75+362336.2 & 
H$\alpha$, H$\beta$, H$\gamma$, H$\delta$, He\,{\sc I}, [O\,{\sc III}] & 
-- & 
Ca\,{\sc I}, Fe\,{\sc I}, Na\,{\sc I}, Ca\,{\sc II} \\

J053008.24+095043.9 & 
H$\alpha$, H$\beta$, H$\gamma$, H$\delta$, He\,{\sc I}, [O\,{\sc III}] & 
-- & 
Ca\,{\sc I}, Fe\,{\sc I}, Na\,{\sc I}, Ca\,{\sc II} \\

J063827.70+102829.1 & 
H$\alpha$, H$\beta$, H$\gamma$, H$\delta$, He\,{\sc I}, [O\,{\sc III}], [Ne\,{\sc III}] & 
-- & 
Ca\,{\sc I}, Fe\,{\sc I}, Na\,{\sc I}, Ca\,{\sc II} \\
\bottomrule
\end{tabular}

\vspace{3mm}
{\footnotesize
\tablecomments{The spectral characteristics of 26 symbiotic star candidates, including their designation, emission features and absorption features.}}
\end{table*}

\end{document}